# Anomalous visualization of sub-2 THz photons on standard silicon CCD and COMS sensors

Mostafa Shalaby[1], Carlo Vicario[1], and Christoph P. Hauri[1,2]

[1]Paul Scherrer Institute, SwissFEL, 5232 Villigen PSI, Switzerland.

[2]École Polytechnique Fédérale de Lausanne, 1015 Lausanne, Switzerland.

*Correspondence to:  most.shalaby@gmail.com, carlo.vicario@psi.ch, and christoph.hauri@psi.ch

**We experimentally show that indirect light-induced electron transitions could lead to THz detection on standard CCD and CMOS sensors, introducing this well-established technological concept to the THz range. Unlike its optical counterpart, we found that the THz sensitivity is nonlinear. We imaged 1-13 THz radiation with photon energy < 2% of the well-established band gap energy threshold. The unprecedented small pitch and large number of pixels uniquely allowed us to visualize the complex propagation of THz radiation, as it focuses down to the physical diffraction limit. Broadband pulses were detectable at a single shot. This opens a whole new field of real time THz imaging at the frame rate of the sensor. The demonstrated experimental finding may redirect THz imager's science to the well-established silicon CCD/CMOS concept which is insensitive to the background thermal environment. This advance will have a great impact on a wide range of science disciplines.**

Photo-induced carrier generation is responsible for the operation of charge-coupled devices (CCDs) [1-2]. In addition to their direct imaging applications in nearly all science disciplines, these sensors come at the core of many advanced imaging methodologies such as photography, holography, diffraction imaging, transmission imaging, tomography, fluorescence spectroscopy etc [3].  CCDs are particularly powerful because they allow for online read-out and single-shot recording at a few micrometer resolution. The typical mode of operation of a CCD sensor depends on an internal photo-electric effect where the incident photons are absorbed in the silicon substrate generating electron-hole pairs across the bandgap. A prerequisite for this process is that the incident photon is energetic enough to overcome the bandgap energy $E_g$. This has restricted the detection to the x-ray, visible and recently to the mid-infrared spectral region. The fundamental limitation on the required photon energy totally prevented the application of CCD sensors to low photon energy/long wavelengths such as terahertz (THz) radiation [4-8]. Here we show that indirect light-matter interaction processes could lead to visualization of low frequency terahertz radiation with photon energy < 2% the energy gap, with high sensitivity and unprecedented resolution. As the presented detector functionality depends on charge transfer and not on thermal induction, our chip allows for measurements insensitive to ambient temperature changes. The presented THz 2D sensor is sensitive to an extremely broad frequency range and is free of any resonant-enhancement structure.

Given the wide applicability of THz radiation [9-16], imaging in this frequency range is crucial [17]. However, similar to other aspects of the THz technology [18-20], THz imaging severely lags behind its counterpart at higher frequency ranges. There have been great efforts to extend the electronic CMOS technology to the THz range, but suffered from high complexity, radiation coupling issues, large pixel size and was limited to the sub-THz range [4-8]. Presently, the state of the art in THz imaging technology is based on a different concept, a bolometer-based detector with a limited number of pixels (320 x 240) and poor spatial resolution (at best 23.5 μm), for example, for applications requiring sub diffraction limit imaging [21]. Furthermore, the response time is slow, the read-out time is long and the thermal sensor is highly sensitive to fluctuation of the ambient temperature as the associated thermal energy is in the range of the THz photon energy to be detected (≈25 meV at 290 K, ≈6 THz). In practice, this gives rise to a detector background radiation which can easily vary during the exposure time. These limitations call for a more robust THz imaging technology. The present work introduces the well-established CCD concept to the THz regime allowing for THz recording on a large size 1360 x 1024 standard silicon chip with a very small pixel size of 4.5 μm.

Our experiment was performed using silicon CCD/CMOS sensors. Typically, the absorbed photons in the photoactive region (silicon layer) generate electron-hole pairs across the band gap. The electrons generated during the exposure time are confined in a potential well in each pixel before the read-out process is launched. For charge separation to take place, the photon quantum needs to carry sufficient energy to overcome the silicon bandgap energy $E_g$ =1.12 eV, corresponding to a wavelength λ of 1.1 μm. In principle, this makes the CCD/CMOS insensitive to radiation with λ > 1.1 μm as the charge-generation region (silicon) becomes transparent. This fundamental limitation prevented the applications of CCD/CMOS sensors to long wavelengths.

The temporal trace of the exciting THz pulse, generated by optical rectification in a DSTMS organic crystal, is single-cycle (Fig. 1a) and the spectrum is centered around 3 THz with components reaching up to 12 THz (Fig. 1b) [25, 26]. In order to eliminate the spectral dependence of our detector, we start by extracting the spectral components in a narrow bandwidth using a 10 THz-centered band pass filter (BPF) with a bandwidth of 1.45 THz (Fig. 1b). We recorded the image of the focused THz beam on the CCD detector (Fig. 1c). We compare it with the image taken with the state-of-the-art bolometer-based camera (from NEC inc.). Two slices of the beam are shown at the focus and 200 μm away. The slices feature multiple spots suggesting that the beam consists of multiple beamlets.

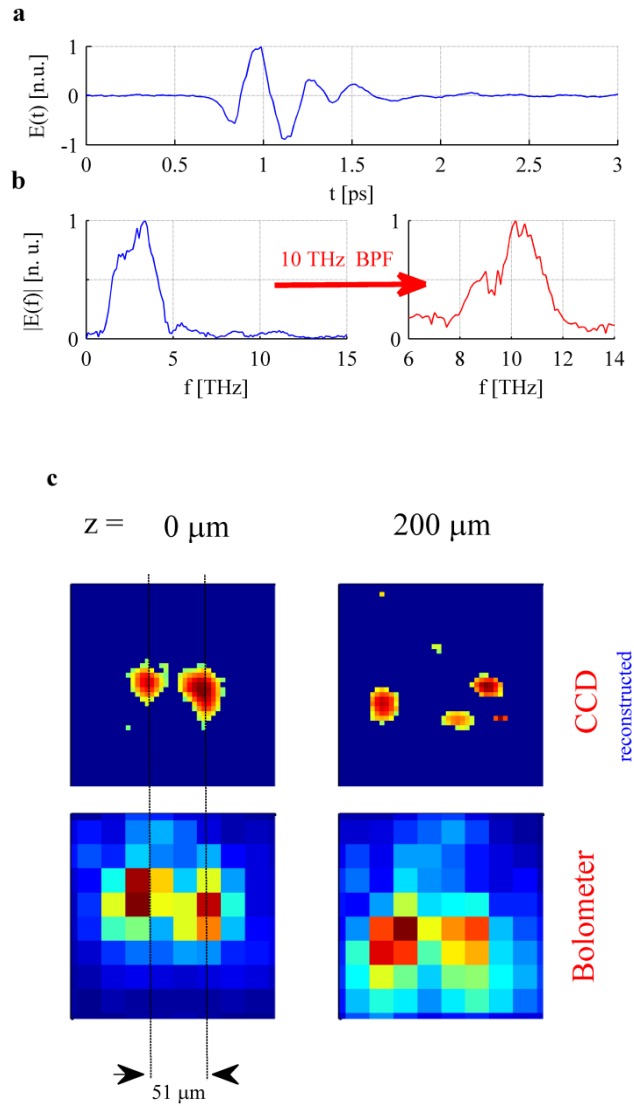

**Figure 1 | Narrowband THz characterization with CCD in comparison with a standard bolometric imager. a,** Time-dependent electric field of the THz pulse retrieved using Air-Biased-Coherent-Detection technique [27]. **b,** the corresponding broadband amplitude spectrum. A narrowband portion centered around 10 THz is obtained by application of a band pass filter (BPF). **c,** The detected images (normalized) on the (4.5 µm-pixel size) CCD obtained with the 10 THz BPF in different slices around the focus with z being the propagation direction. The corresponding real images reconstructed (as discussed later). The beam consists of several smaller beams that focus in different planes. The images obtained with a (23.5 µm-pixel size) bolometric imager are shown. All images are plotted at the same scale on dimension of 200 µm x 200 µm. In the focal plane (z = 0 µm), two spots are visible with a separation of 51 µm. While the pixel size is smaller than 51 µm in both cameras and so the two spots were obvious in both images, the high resolving image from the CCD provides a significantly more detailed and accurate beam profile.

As depicted, the THz beam at the focus consists of several individual beamlets. Although both sensors show this feature, only a diffuse image is observed with the bolometric sensor. In contrast, highly detailed information is provided by the high-resolving CCD sensor. For example, in the assumed focal plane, the two spots are resolved using the CCD to be separated by 51 µm which corresponds to nearly two pixels from the bolometer array, thus incapable of spatial sampling of the intensity profile. In contrast, the CCD was capable of easily resolving features which are a fraction of the wavelength, down to $\lambda/6.5$. The shown images are rescaled in intensity using a calibration described later. Furthermore, the CCD sensor offers a signal to noise ratio (SNR) better than 64 for a single shot exposure and a SNR of 6400 for the maximum exposure time of 1 second with the laser system running at 100 Hz. The maximum total energy shined on the detector was. In our experiment, we carried out extensive tests to rule out any contribution from the residual optical pumping beam reaching the detector. This includes.

In order to get insight on the spatial evolution of the full THz pulse bandwidth, we show in Figure 2 the beam profile images recorded with both sensors. Only with the CCD is the complex propagation behavior of the multiple beamlets visible. They are shown to have different divergences, sizes, and focal planes around the assumed focus. In comparison, the large pixel size of the bolometric sensor smears these features out.

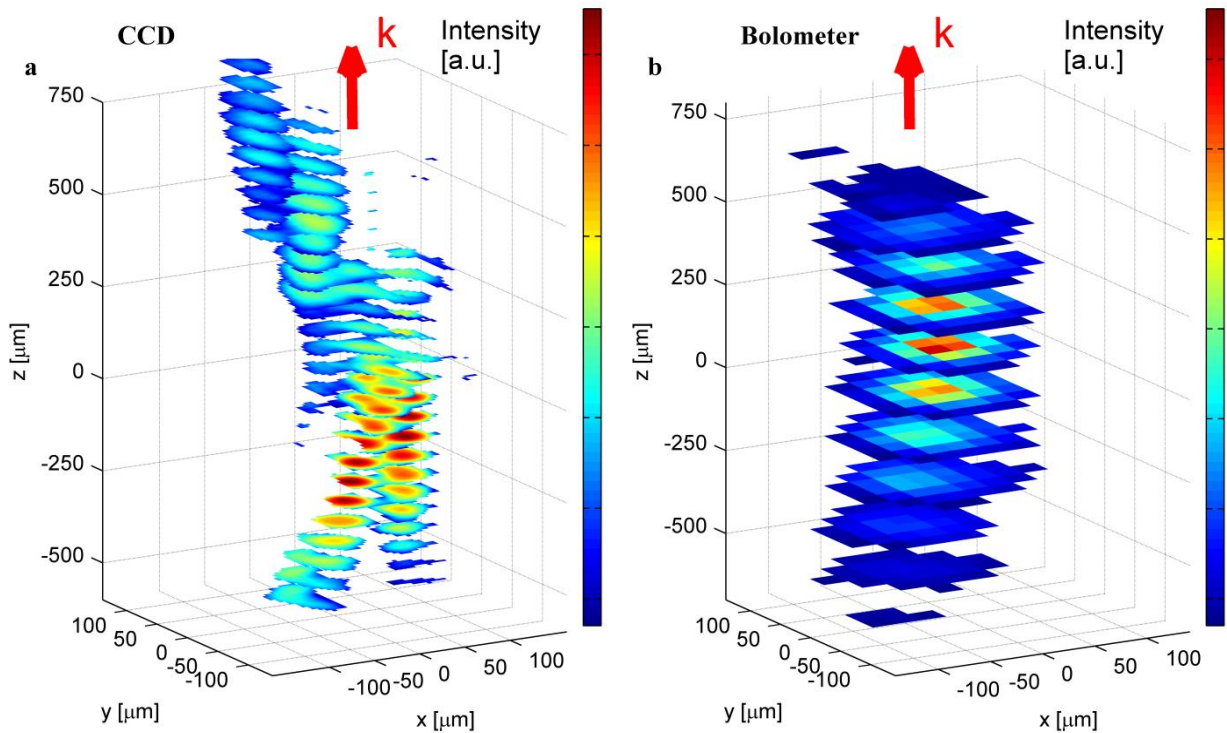

**Figure 2 | Three-dimensional spatial evolution of the THz beam across the beam waist.** Images were retrieved using **a,** CCD and **b,** bolometric sensors. In the case of CCD, multiple beams with different divergences, sizes, and focal planes are clearly visualized. The corresponding images taken from the bolometric detector barely show this detailed-rich structure of the beam due to poor resolution resulting from the large pixel size. The bolometric sensor has a very strong frequency-dependent sensitivity [28], but it is still sensitive to the low frequency components.

We measured the peak intensity of the THz image on the CCD for different exposure times (pulse count) and verified the expected response (Fig. 3a). The minimum exposure time was 10 ms corresponding to a single THz pulse at the laser repetition rate of 100 Hz. This verification allows us to extract the dependence of the measured intensity of the CCD on the THz peak intensity/pulse energy.

It is likely that both processes contribute sequentially. Nevertheless, our band gap and operating room temperature strongly differentiates the process under investigation here from previous studies on THz-induced nonlinearities in semiconductors.

The frequency-dependent sensitivity of the CCD was studied by employing a set of available THz filters including a bandpass filter (BPF) centered at 10 THz and two low-pass filters with cutoff freuencies at 18 THz (LP1) and 9 THz (LP2), repectively. As illustrated in Figure 3a, dependence for these frequency regions was found, such as for the BPF (green), the LP1 (red) and for LP2 (blue). From these measurements it becomes clear that the CCD shows a frequency-dependent sensitivity which is larger for.

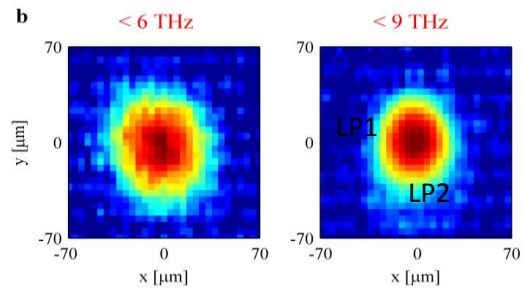

**Figure 4 | Spectral sensitivity of the CCD camera. a,** The dependence of the image intensity on the THz pulse energy in the case of 10 THz band pass filter BPF (green), 18 THz low pass filter LP1 (red), and 9 THz low pass filter LP2 (blue). The multiplier denotes the required expansion factor of the horizontal axis. **b**, Terahertz images reconstructed using CCD in the sub-6 THz and sub-9 THz ranges with the unprecedented resolution of 4.5 µm.

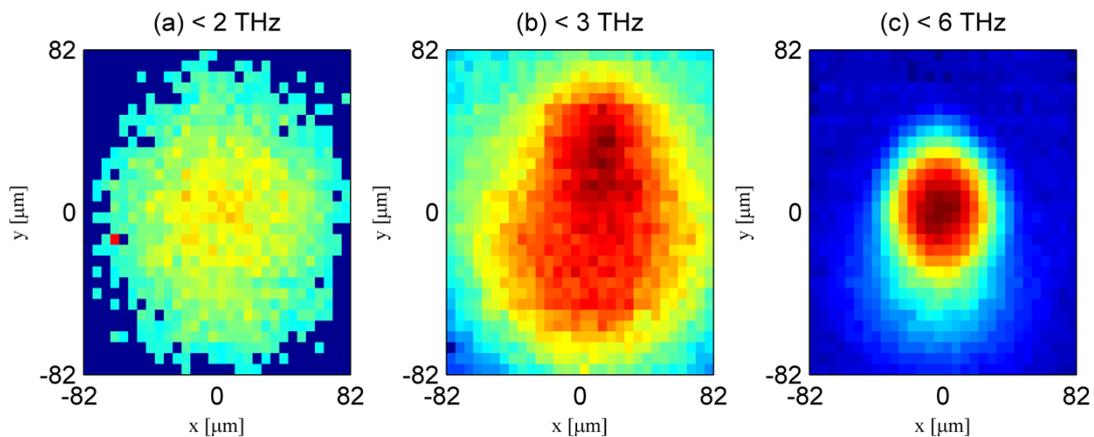

**Figure 5 | Spectral sensitivity of the CMOS camera.** Terahertz images reconstructed using CMOS sensor in the sub-2 THz, sub-3 THz and sub-6 THz ranges with the unprecedented resolution of µm. These are obtained at different source conditions from those in Figure. 4

In conclusion, we have introduced the well-established CCD/CMOS concept to the technologically lagging THz imaging. We have shown that indirect radiation mechanism could lead to broadband low frequency (1-13 THz) imaging on standard silicon CCD detectors. The exceptionally small pixel size allowed us to see the detailed and complex evolution of a THz beam profile in space. The small pixel size (4.5 µm), single shot capabilities, and the large number of pixels open an avenue for novel THz imaging applications such as holography, and for advanced methodology such as THz wavefront sensors. Many disciplines including fundamental science, medicine and national security would greatly benefit from these novel applications.